\documentclass[conference]{IEEEtran}

\usepackage{fancyhdr, epsfig, epsf, amsthm, amsmath, amssymb, amsfonts, subfigure, color}
\usepackage{threeparttable}
\usepackage[noadjust]{cite}
\usepackage{dsfont}
\usepackage{enumerate}
\usepackage{comment}
\usepackage{multirow}
\usepackage{bigstrut}

\newtheorem{lemma}{Lemma}
\newtheorem{corollary}{Corollary}

\newtheorem{proposition}{Proposition}
\newtheorem{remark}{Remark}

\def\qed{$\blacksquare$}

\def\endproof{\hfill \qed}

\def\E{\mathsf{E}}

\def\SIR{\mathsf{SIR}}

\def\l{\left}
\def\r{\right}
\def\({\left(}
\def\){\right)}

\def\[{\left[}
\def\]{\right]}

\def\SIRn{\widetilde{\SIR}}

\def\ufr{\mathsf{FR}_u}
\def\bfr{\mathsf{FR}_b}
\def\ufrmath{\emph{\text{$\ufr$}}}
\def\bfrmath{\emph{\text{$\bfr$}}}

\def\papertitle{Revisiting Frequency Reuse towards Supporting Ultra-Reliable Ubiquitous-Rate Communication}
\IEEEoverridecommandlockouts

\begin{document}
\title{ \fontsize{24}{28}\selectfont  \papertitle}

\author{Jihong~Park, Dong~Min~Kim, Petar~Popovski, and Seong-Lyun~Kim\IEEEauthorrefmark{2}
\thanks{J.~Park, D.~M.~Kim and P.~Popovski are with Department of Electronic Systems, Aalborg University, Denmark (email: \{jihong, dmk, petarp\}@es.aau.dk). }
\thanks{\IEEEauthorrefmark{2}S.-L.~Kim is with School of EEE, Yonsei University, Seoul, Korea (email: slkim@yonsei.ac.kr). }
\thanks{This work has been supported in part by the Danish Ministry of Higher Education and Science (EliteForsk Award, Grant Nr. 5137-00073B), and partly by the National Research Foundation of Korea (NRF) grant funded by the Korea government (MSIP) (NRF-2014R1A2A1A11053234).}
}
\maketitle \thispagestyle{empty}

\begin{abstract}
One of the goals of 5G wireless systems stated by the NGMN alliance is to provide moderate rates (50+ Mbps) everywhere and with very high reliability. We term this service  \emph{Ultra-Reliable Ubiquitous-Rate Communication (UR2C)}. This paper investigates the role of frequency reuse in supporting UR2C in the downlink. To this end, two frequency reuse schemes are considered: \emph{user-specific} frequency reuse ($\ufr$) and \emph{BS-specific} frequency reuse ($\bfr$). For a given unit frequency channel, $\ufr$ reduces the number of serving user equipments (UEs), whereas $\bfr$ directly decreases the number of interfering base stations (BSs). This increases the distance from the interfering BSs and the signal-to-interference ratio ($\SIR$) attains ultra-reliability, e.g. $99\%$ $\SIR$ coverage at a randomly picked UE. The ultra-reliability is, however, achieved at the cost of the reduced frequency allocation,
which may degrade overall downlink rate. To fairly capture this reliability-rate
tradeoff, we propose \emph{ubiquitous rate} defined as the maximum downlink rate whose required $\SIR$ can be achieved with ultra-reliability. By using stochastic geometry, we derive closed-form ubiquitous rate as well as the optimal frequency reuse rules for UR2C.
\end{abstract}

\section{Motivation and Contribution}

Two extremes of 5G wireless system designs are pursuing very high data rate, i.e.
enhanced mobile broadband (eMBB), and enabling ultra-reliable low-latency
communication (URLLC) with very low data rate \cite{NGMNKPI:16,Qualcomm5G:16,PetarURC:14}. This paper explores a missing piece between them, \emph{Ultra-Reliable Ubiquitous-Rate Communication} (UR2C), answering to the question how much data rate can stably be achieved everywhere. For example, the NGMN Alliance suggests that a rate of at least $50$ Mbps should be supported practically everywhere \cite{NGMNKPI:16}. In this paper ``practically'' means that a randomly picked user equipment (UE) can ubiquitously attain a predefined data rate with $99$\% reliability, which is higher than the commonly considered $95$\% reliability of the wireless coverage \cite{NGMNKPI:16,PetarURC:14}.

An important enabler of UR2C is ultra-dense base station (BS) deployment where BS/UE density ratio \emph{per communication resource} exceeds $1$ \cite{JHParkTWC:15}. A larger density of BSs improves signal-to-interference ratio ($\SIR$) since it increases the number of non-interfering BSs that have no serving UE within their cells. This makes the desired received signal power grow faster than the interference increase when deploying more BSs.

\begin{figure}
	\centering
	\includegraphics[width=8.5cm]{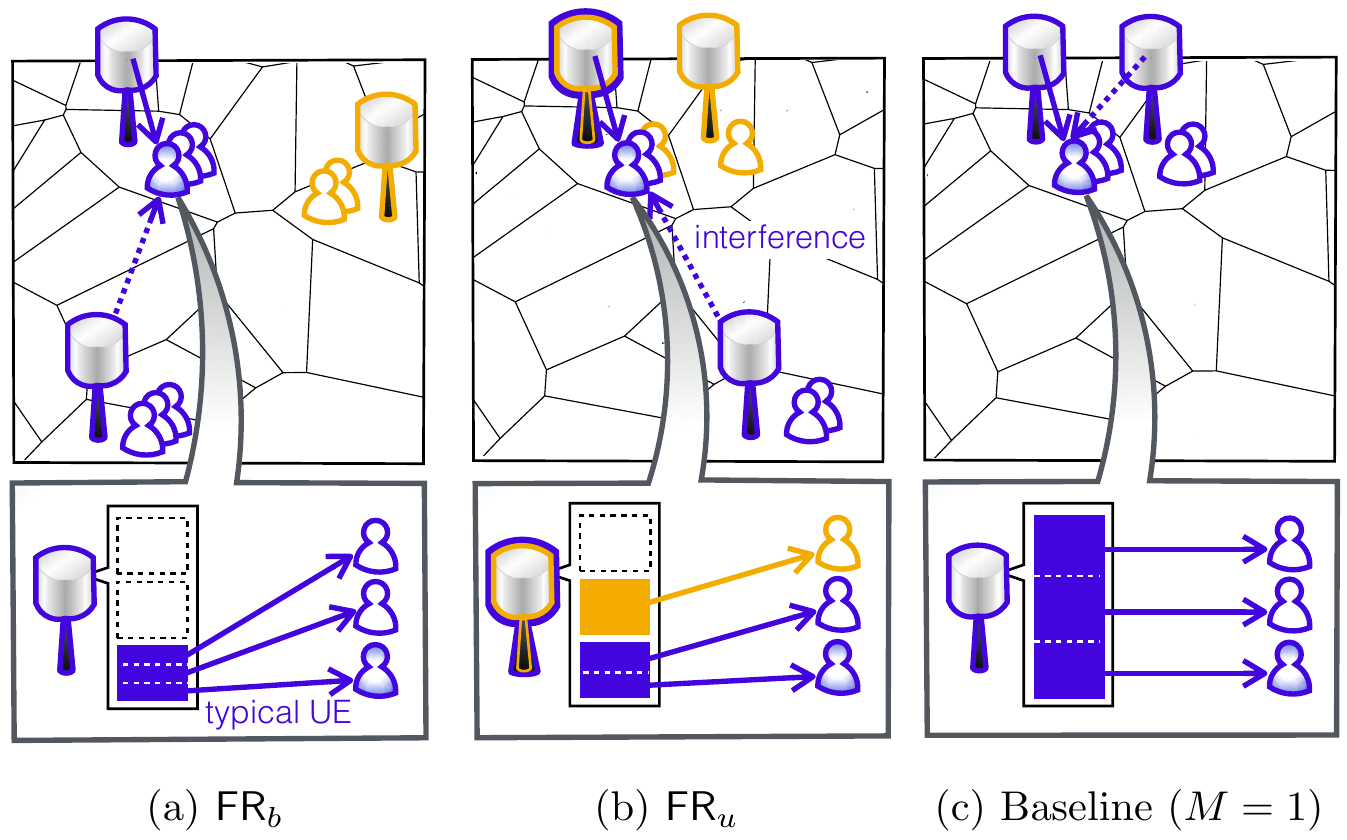}
	\caption{Illustrations of $\ufr$ and $\bfr$ when the number of channels $M=3$: (a) In $\bfr$, a BS randomly selects a single channel for all associated UEs; (b) In $\ufr$, a BS randomly assigns a single channel to each associated UE. At a typical UE (filled blue), both $\bfr$ and $\ufr$ thereby increase interfering BS distance, compared to (c) the baseline's when $M=1$.}
\end{figure}

The effect of ultra-densification can also be achieved by dividing the total bandwidth into frequency channels. It makes interfering BSs located farther away, yielding $\SIR$ improvement. This motivated us to reconsider the problem of \emph{frequency reuse}.
In a downlink scenario, we fine-tune frequency reuse to make it applicable for a randomly selected UE, i.e. a \emph{typical UE}. Thereby we aim at proposing an optimal frequency reuse scheme that achieves our target ultra-reliability, $99\%$ $\SIR$ coverage. To this end, we suggest two frequency reuse techniques: \emph{BS-specific} frequency reuse ($\bfr$) and \emph{user-specific} frequency reuse ($\ufr$). When the entire frequency bandwidth is divided into $M$ number of channels, each BS in $\bfr$ uses only a single channel that it selects randomly (see Fig. 1-a). In $\ufr$, a BS randomly assigns a single channel to each associated UE (Fig. 1-b). If all the assigned channels to UEs are not identical, its corresponding BS utilizes multiple number of channels. Nevertheless, average amount of $\ufr$'s channel use is still less than a baseline without frequency reuse, i.e. $M=1$ (Fig. 1-c). As every scheme with $M$ larger than $1$, both $\bfr$ and $\ufr$ introduce loss in the data rate as they are only partially using the bandwidth.

To make the proposed schemes applicable not only for a cell-edge UE but also for a typical UE, it is therefore important to examine how much data rate degrades in return for reliability improvement. For this purpose, we propose \emph{ubiquitous rate} $\mathcal{R}_{\eta}$
defined as the maximum ergodic capacity that guarantees a target $\eta$ $\SIR$ coverage
probability for a given $\SIR$ threshold $t$. For instance, $\mathcal{R}_{0.99}$ for unit resource allocation is $0.99\cdot \log_2(1+ t)$~bps, where $t$ is given
by taking the inverse function of $\SIR$ coverage probability to the target
\emph{reliability constraint equation} $\Pr(\SIR\geq t) = 0.99$.

In this respect, we utilize stochastic geometry, and provide the closed-form downlink ubiquitous rate at a typical UE. Its derivation follows in part from preceding works that provide closed-form downlink $\SIR$ coverage at a typical UE in a form of a hypergeometric function \cite{Andrews:2011bg,Haenggi:ISIT14}. However, inverting a hypergeometric function to solve the reliable constraint equation is not analytically viable. We instead propose novel closed-form $\SIR$ bounds by using a hypergeometric function transformation technique, and thereby derive the desired ubiquitous rate in a closed form. Based on this result, we optimize $\bfr$ and $\ufr$ with respect to $M$, and investigate how to achieve the UR2C's target $50$ Mbps ubiquitous rate with $99\%$ reliability under different deployment density scenarios.

The contributions of this paper are summarized as follows.
\begin{itemize}
\item Optimized frequency reuse designs for UR2C are proposed (see \textbf{Proposition 3} with Figs. 3 and 4).
\item Closed-form ubiquitous rates and $\SIR$ reliabilities under the frequency reuse schemes are derived (\textbf{Proposition 2}).
\item Closed-form $\SIR$ reliability approximation for UR2C as well as closed-form $\SIR$ reliability bounds are provided (\textbf{Lemmas 2} and \textbf{3}).
\end{itemize}

\section{System Model}
In a downlink cellular network, BSs are uniformly distributed with density $\lambda_b$,
resulting in a homogeneous Poisson point process (PPP). Independently, UEs are also
uniformly distributed with density $\lambda$, leading to another homogeneous PPP with density $\lambda$. Each UE associates with the nearest BS, forming a Poisson-Voronoi
tessellation \cite{HaenggiSG}.

The entire frequency bandwidth $W$ is divided into $M_i$  channels, where hereafter
the subscript $i\in\{b, u, o\}$, respectively, denotes $\bfr$, $\ufr$, and a baseline model. In $\bfr$, each BS randomly selects a single frequency channel, and serves its associated UEs only via the selected channel. In $\ufr$, a BS randomly assigns a single channel to each of its associated UE, and serves them respectively through the UEs' assigned channels. In the baseline, BSs neglect frequency reuse, and serve their associated UEs by using the entire bandwidth, i.e. $M_o=1$. For all cases, multiple UEs using the same bandwidth are served by the BS through an equal TDMA allocation.

A BS occupies each channel when at least a single UE is served via the channel; otherwise, the channel remains void for that BS and the BS is considered inactive. For $\lambda$ density of UEs served by a BS via a single channel, the BS's corresponding \emph{channel occupancy probability} $p_c(\lambda)$ is given as $p_c(\lambda) \approx 1-\(1+\lambda / \[3.5 \lambda_b \] \)^{-3.5}$ \cite{Yu2011}, which is a monotone increasing (or decreasing) function of $\lambda$ (or $\lambda_b$). In sparse networks $\lambda_b/\lambda\rightarrow 0$, $p_c(\lambda)$ becomes $1$, while for ultra-dense networks $\lambda_b/\lambda \rightarrow\infty$, $p_c(\lambda)$ decreases towards $0$ at the rate $\lambda/\lambda_b$ \cite{JHParkTWC:15}.

For each non-void channel, a BS allocates unity power, and transmits signals. The transmitted signals experience path-loss attenuation with the
exponent $\alpha>2$ and Rayleigh fading with unit mean. Noise power is assumed to be much
smaller than interference, and is thus neglected as in \cite{Andrews:2011bg,Haenggi:ISIT14,JHParkTWC:15}.

\begin{table}[!t]
\renewcommand{\arraystretch}{1.2}
\caption{List of Notations}
\label{Table:Notations}
\centering
\begin{tabular}{ r  l }
\hline
\hspace{-10pt}\bf{Notation} &\hspace{-5pt} \bf{Meaning}\\
\hspace{-10pt}$\textsf{FR}_b$, $\textsf{FR}_u$ &\hspace{-5pt} BS/User-centric frequency reuse\\
\hspace{-10pt}$W$ &\hspace{-5pt} Total frequency bandwidth\\
\hspace{-10pt}$M_i$ &\hspace{-5pt} \# Channels where $i\in\{b, u, o\}$ resp. for $\bfr$, $\ufr$ and baseline\\
\hspace{-10pt}$\SIR(M_i)$ &\hspace{-5pt} $\SIR$ under $M_i$ number of channels\\

\hspace{-10pt}$N_i$ &\hspace{-5pt} \# UEs sharing the same BS and channel with a typical UE\\
\hspace{-10pt}$\mathcal{P}_t(M_i)$ &\hspace{-5pt} $\SIR$ reliability with $\SIR$ threshold $t$\\
\hspace{-10pt}$\eta$ &\hspace{-5pt} Target $\SIR$ reliability; e.g. 0.99 for $99\%$ $\SIR$ reliability \\
\hspace{-10pt}$R_{t}(M_i)$ &\hspace{-5pt} Average rate, given as $W/M_i \cdot \E\[1/N_i\] \cdot \log(1 + t)$ \\
\hspace{-10pt}$\mathcal{R}_\eta(M_i)$ &\hspace{-5pt} Ubiquitous rate, maximum avgerage rate s.t. $\mathcal{P}_t(M_i)\geq \eta$\\
\hspace{-10pt}$\lambda_b, \lambda$&\hspace{-5pt} BS and UE densities \\
\hspace{-10pt}$p_c(\lambda)$ &\hspace{-5pt} Channel occupancy probability for serving UE density $\lambda$\\
\hline
\end{tabular}
\end{table}

\section{Problem Formulation: Ubiquitous Rate Maximization}
At a typical UE, the reliability of $\SIR$ $\mathcal{P}_t(M_i)$ when there are $M_i$ channels is defined as:
\begin{align}\label{Eq:Reliability}
\mathcal{P}_t(M_i) &:= \Pr\big(\SIR(M_i) \geq t\big)
\end{align}
where $\SIR(M_i)$ is the $\SIR$ with $M_i$, elaborated in Section IV. Note that $\mathcal{P}_t(M_i)$ is a monotone increasing function of $M_i$. As $M_i$ increases, interfering BSs become located farther away, which thereby improves $\SIR(M_i)$ and so does $\mathcal{P}_t(M_i)$.

Increasing $M_i$, however, may decrease data rate since it reduces per-UE resource allocation at least by $M_i$. Precisely, the typical UE's downlink rate $R_{t}(M_i)$ is given as
\begin{align}\label{Eq:Rate}
R_t(M_i) &:= \underbrace{\frac{W}{M_i} \cdot \E\[\frac{1}{N_i}\]}_{\text{resource allocation}}  \cdot \underbrace{\log_2(1 + t)}_{\text{spectral efficiency}}
\end{align}
where $N_i$ is the number of UEs served by the same BS via the same channel as the typical UE's.

The first term of resource allocation $W/M_i$ in \eqref{Eq:Rate} indicates the maximum resource allocation per UE when a BS serves only a single UE, which is reduced by $1/M_i$. The actual resource allocation to a typical UE is $\E[1/N_i]$ portion of $W/M_i$ since the maximum amount is equally allocated to the multiple UEs sharing the same channel and BS with the typical UE's. According to \cite{Yu2011}, the reciprocal of $\E[1/N_i]$ can be calculated by multiplying (i) the density of common UEs using the same channel as the typical UE's and (ii) average cell size of the BSs serving such common UEs. In $\bfr$, each BS utilizes only a single channel, so the entire UEs with density $\lambda$ become the common UEs. Next, the density of the BSs serving the common UEs is $p_c(\lambda)\lambda_b$. The average size of cells having the common UEs then becomes $1/\(p_c(\lambda)\lambda_b\)$ because the average cell size of BSs is the reciprocal of the BS density \cite{StoyanBook:StochasticGeometry:1995}. As a result, $\E[1/N_b]=p_c(\lambda)\lambda_b/\lambda$. In the baseline, common UE density is also $\lambda$, so the same calculation applies, i.e. $\E[1/N_o]=\E[1/N_b]$. In $\ufr$, common UE density is reduced by $1/M_u$, and this enlarges the average cell size by replacing $\lambda$ with $\lambda/M_u$, as summarized in Table~II.

\begin{table}[!t]
\renewcommand{\arraystretch}{1.2}
\caption{List of Resource Allocation and Interferer Density}
\label{Tab:SRNRValues}
\begin{center}
\begin{tabular}{r|l|l|l}
\hline
\multirow{2}{.9cm}{\textbf{Scheme}}& \multicolumn{2}{p{4cm}|}{\centering \textbf{Resource Allocation}} & \textbf{Interferer Density}\\
 & \multicolumn{1}{c|}{$W/M_i$} & \multicolumn{1}{c|}{$\E[1/N_i]$} & \multicolumn{1}{c}{$\lambda_i$} \\ \hline
Baseline & $\;W$  & $p_c(\lambda)\lambda_b/\lambda$ & $  p_c\(\lambda\) \lambda_b$\\
\;\;\;\;\;\;$\bfr$& $\;W/M_b$  & $p_c(\lambda)\lambda_b/\lambda$  & $ p_c\(\lambda\) \lambda_b/M_b$ \\
\;\;\;\;\;\;$\ufr$& $\;W/M_u$  & $p_c(\lambda/M_u)\lambda_b/(\lambda/M_u)$  & $ p_c\(\lambda/M_u\) \lambda_b$ \\
\hline
\end{tabular}
\end{center}
\end{table}


Consequently, in return for resource allocation reduction as $M_i$ increases, it improves $\SIR$ reliability in \eqref{Eq:Reliability}, and thereby increases spectral efficiency in \eqref{Eq:Rate} for a fixed target reliability. To achieve UR2C, we  capture this reliability-rate tradeoff between \eqref{Eq:Reliability} and \eqref{Eq:Rate}, and formulate our optimization problem.
\begin{align}
\vspace{-10pt}\texttt{(P0)}\;  \underset{M_i}{\texttt{maximize}} &\; R_t(M_i) \nonumber\\
\texttt{s.t.} &\; \mathcal{P}_t(M_i) \geq \eta \nonumber
\end{align}

It is noted that $R_t(M_i)$ and $\mathcal{P}_t(M_i)$ respectively are monotone increasing and decreasing functions of the $\SIR$ threshold $t$. Therefore $R_t(M_i)$ is maximized when $\mathcal{P}_t(M_i)=\eta$ holds, which is the \emph{reliability constraint equation}. In this respect, we define \emph{ubiquitous rate} $\mathcal{R}_\eta(M_i)$ as the maximum rate
guaranteeing $\mathcal{P}_t(M_i) \geq \eta$. 

Our objective is to derive $\mathcal{R}_\eta(M_i)$ for each frequency reuse scenario. To this end, we first derive a closed-form $\mathcal{P}_t(M_i)$ approximation (Lemma 3), and thereby easily solve the reliability constraint equation with respect to $t$. For this given optimal $t$, we tractably optimize $\mathcal{R}_\eta(M_i)$ with respect to $M_i$ under an ultra-reliable regime, i.e. $\eta\approx 1$.

\section{Closed-Form Reliability via Normalized $\SIR$}

This section derives closed-form $\SIR$ reliability approximation for UR2C (Lemma 3). This enables closed-form ubiquitous rate expressions in the next section (Proposition 2). To this end, we consider the following two techniques.

\textbf{Technique \#1. $\SIR$ normalization}. At a typical UE, let $S\(\lambda_b\)$ denote its received desired signal power when associating with the nearest BS out of all BSs with density $\lambda_b$. Similarly, $I_r\(\lambda_i\)$
denotes aggregate interference where $\lambda_{i}$ is \emph{interferer density} and $r$ the typical UE's association distance. In the baseline utilizing the entire bandwidth, all the BSs that have at least a single UE perform downlink transmissions interfering with the typical UE. Therefore, $\lambda_o = p_c(\lambda)\lambda_b$. In $\bfr$, $1/M_b$ portion of such non-void BSs become interferers, and thus $\lambda_b = p_c(\lambda)\lambda_b/M_b$. In $\ufr$, only the BSs having at least a single UE that is served through the same channel as the typical UE's with probability $1/M_u$ become interferers. It leads to $\lambda = p_c(\lambda/M_u)\lambda_b$, as summarized in Table II.

Next, we utilize a transformation between BS density $\lambda_b$ and received signal power $S\(\lambda_b\)$ (or $ \lambda_i$ and $I_r(\lambda_i)$). According to mapping theorem \cite{HaenggiSG}, $P$ times transmission power increase for all BSs is almost surely ($a.s.$) identical to $P^{\frac{2}{\alpha}}$ times BS densification without their power increase, from the typical UE's received signal power point of view. By exploiting this relationship, we can transform any $\SIR(M_i)=S\(\lambda_b\)/I_r\(\lambda_i\)$ into its normalized value $\SIRn = S(1)/I_r(1)$ specified in the following lemma.

\begin{lemma} \emph{(SIR normalization, Corollary 2.35 in \cite{HaenggiSG})} $\SIR(M_i)$ and its normalized $\SIRn$ satisfy the following relationship.

\vspace{-10pt}\small\begin{align}
\SIR(M_i) = \(\frac{\lambda_b}{\lambda_i}\)^\frac{\alpha}{2}  \SIRn \quad a.s.
\end{align}\normalsize

\end{lemma}
\noindent This allows us to focus only on $\lambda_i$ when examining the proposed scheme's impact on $\SIR$.

\textbf{Technique \#2. Hypergeometric function transformation}.
When $\lambda_i=\lambda_b$, it has been known that the baseline model's downlink $\SIR$
coverage at a typical UE is the reciprocal of a Gauss-hypergeometric function
\cite{Andrews:2011bg,Haenggi:ISIT14}, given as

\vspace{-10pt}\small\begin{align}\label{Eq:OldCoverage}
\mathcal{P}_t(M_i) &= \;_2F_1\(1, -\frac{2}{\alpha}; 1-\frac{2}{\alpha}; -t \)^{-1}
\end{align}\normalsize
where a Gauss hypergeometric function $\;_2F_1(a, b;c;z):= \sum_{n=0}^\infty \frac{(a)_n \(b\)_n }{\(c\)_n} \frac{z^{n}}{n!}$ and Pochhammer symbol $(x)_n$ is defined as $\Gamma(x+n) / \Gamma(x)$ for the Gamma function $\Gamma(x+1):=x!$.

Applying Lemma 1 generalizes this result for an arbitrary $\lambda_i>0$ by using Lemma 1.

\vspace{-10pt}\small\begin{align}
\mathcal{P}_t(M_i) &= \Pr\(\SIRn \geq  \[\frac{\lambda_i}{\lambda_b}\]^\frac{\alpha}{2} t \)\\
&= \;_2F_1\(1, -\frac{2}{\alpha}; 1-\frac{2}{\alpha}; -\[\frac{\lambda_i}{\lambda_b}\]^\frac{\alpha}{2} t \)^{-1}  \label{Eq:PfLemma2_pre}
\end{align}\normalsize
The last step follows from \eqref{Eq:OldCoverage}, which is also applicable for $\SIRn$. Setting $\lambda_i=\lambda_b=1$ while replacing $t$ with $(\lambda_i/\lambda_b)^\frac{\alpha}{2}t$ yield such a result.

The result \eqref{Eq:PfLemma2_pre} is, however, still not applicable for analytically solving the reliability constraint equation $\mathcal{P}_t(M_i) =\eta$ since the inverse function of \eqref{Eq:PfLemma2_pre} is unknown. We detour this problem by utilizing Pfaff's Gauss-hypergeometric function transformation \cite{GEAndrews:Book}, specified in the following lemma.

\begin{lemma} \emph{(Reliability Bounds)} $\SIR$ reliability $\mathcal{P}_t(M_i)$ is upper and lower bounded as follows.
{\small\begin{align}
\frac{\alpha}{2\pi \csc\(\frac{2 \pi}{\alpha}\)}\( 1 + \[\frac{\lambda_i}{\lambda_b}\]^\frac{\alpha}{2} t \)^{-\frac{2}{\alpha}} \hspace{-5pt}\leq \mathcal{P}_t(M_i) \leq  \( 1 + \[\frac{\lambda_i}{\lambda_b}\]^\frac{\alpha}{2} t \)^{-\frac{2}{\alpha}}
\end{align}}\normalsize

\begin{proof}\emph{ Consider Pfaff's transformation \cite{GEAndrews:Book} given as follows.
\\
\vspace{-10pt}\small\begin{align}
\;_2F_1\(c-a, b; c; \[1-1/z\]^{-1}\) &= (1-z)^{-b} \;_2F_1(a, b;c;z)
\end{align}\normalsize
Setting $x = t(\lambda_I/\lambda_S)^\frac{\alpha}{2}$ and applying $a=b=-\frac{2}{\alpha}$, $c=1-\frac{2}{\alpha}$, and $z=\(1 + 1/x\)^{-1}$ yields $\mathcal{P}_t(M_i) = c_0 \(1 + x \)^{-\frac{2}{\alpha}}  $
where $c_0 := _2F_1\(-\frac{2}{\alpha}, -\frac{2}{\alpha}; 1-\frac{2}{\alpha} ; \frac{1}{1+x} \)$ is a monotonically decreasing function of $x$, having maximum $\frac{2\pi}{\alpha} \csc\(\frac{2 \pi}{\alpha}\)$ and minimum $1$ respectively when $x$ approaches $0$ and $\infty$. Applying these inequalities provides the desired result.}
\end{proof}
\end{lemma}

As shown in Fig.~2, the lower bound in Lemma 2 is only accurate for low $\SIR$ reliability when $t\rightarrow \infty$, so is not applicable for UR2C design. On the contrary, the upper bound has potential for the use in an ultra-reliable regime. This motivates us to further improve the upper bound.

With this end, we modify the upper bound so that it can asymptotically converge to the exact curve in the ultra-reliable regime, i.e. $t\rightarrow 0$. This only requires replacing the outermost exponent $2/\alpha$ with $2/(\alpha-2)$, leading to another lower bound.


\begin{figure}\label{Fig:ReliabilityBound}
	\centering
	\hspace{-5pt}\includegraphics[width=8.6cm]{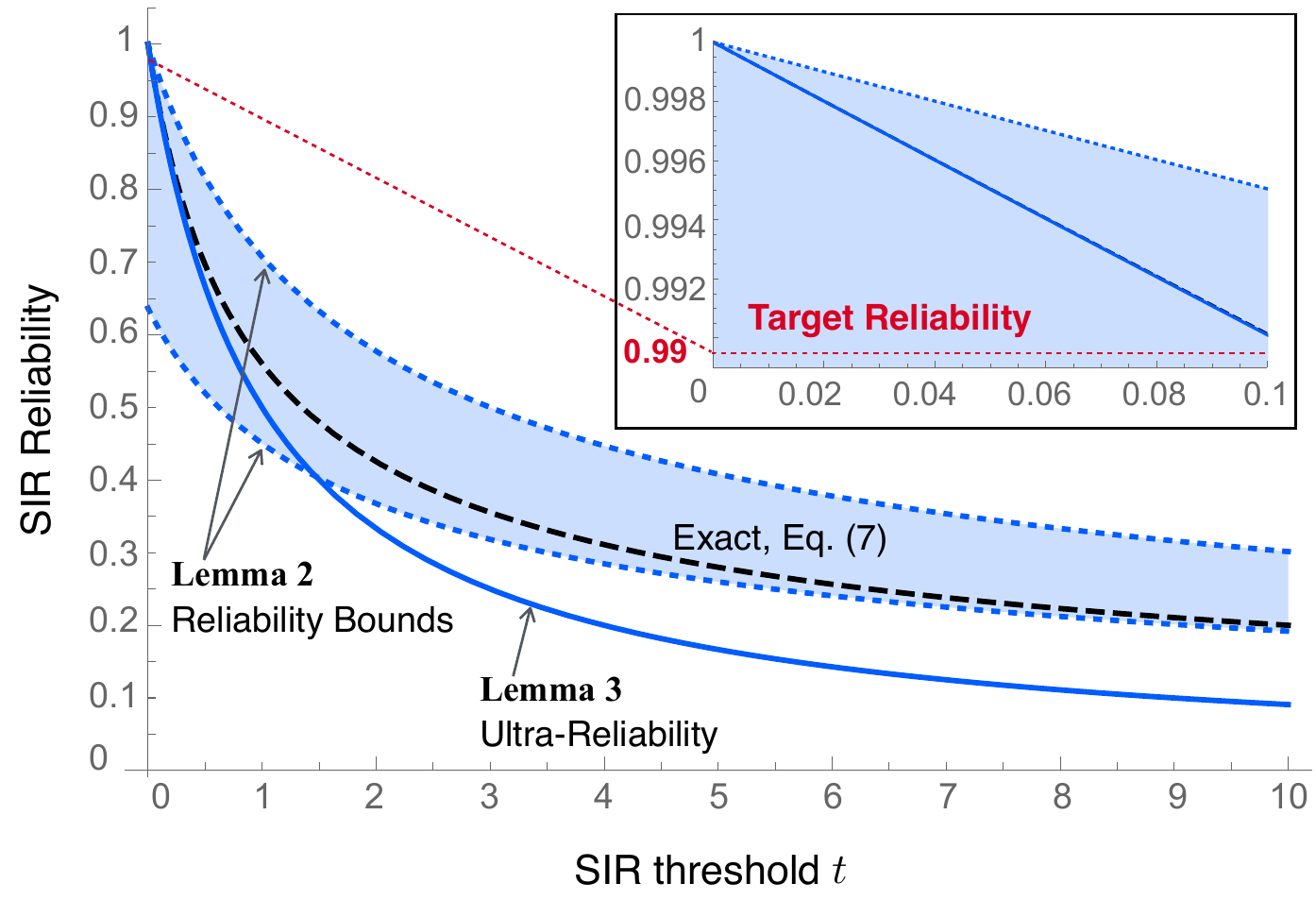}
	\caption{Reliability bounds in Lemmas 2 and 3 ($\lambda_i=\lambda_b$, $\alpha=4$).}
\end{figure}

\begin{lemma} \emph{(Closed-Form Ultra-Reliability)} $\SIR$ reliability $\mathcal{P}_t(M_i)$ is lower bounded as follows.

\vspace{-10pt}\small\begin{align} \mathcal{P}_t(M_i) &\geq  \( 1 +
\[\frac{\lambda_i}{\lambda_b}\]^\frac{\alpha}{2} t \)^{-\frac{2}{\alpha-2}}
\end{align}\normalsize
where the equality holds when $\(\lambda_i/\lambda_b\)^{\frac{\alpha}{2}}t\rightarrow 0$

\begin{proof}
See Appendix.
\end{proof}
\end{lemma}

 As Fig.~2 shows, Lemma 3 lower bound is accurate in an ultra-reliable regime. In the following we therefore utilize this as an approximation, i.e. $\mathcal{P}_t(M_i)\approx \( 1 +
\[\lambda_i /\lambda_b \]^\frac{\alpha}{2} t \)^{-\frac{2}{\alpha-2}}$.

\section{Ubiquitous rate with Frequency Reuse}
In the first subsection, closed-form ubiquitous rates for $\bfr$ and $\bfr$ are derived. Based on this result, $\bfr$ and $\ufr$ are optimized in the second subsection.

\subsection{Closed-Form Ubiquitous Rate}
Applying Lemma 3 with $\lambda_i$ summarized in Table II yields the closed-form reliabilities for $\bfr$, $\ufr$, and the baseline.
\begin{proposition} \emph{(Reliability)} For $\eta \approx 1$, the reliabilities of $\bfrmath$, $\ufrmath$, and the baseline for $\SIR$ threshold $t$ are given as follows.

\vspace{-10pt}
\small\begin{align}
\bfrmath: \quad \mathcal{P}_t (M_b) &\approx \[ 1 + \(\frac{p_c\(\lambda\)}{M_b}\)^{\frac{\alpha}{2}}t \]^{-\frac{2}{\alpha-2}}\\
\ufrmath: \quad \mathcal{P}_t (M_u)  &\approx \[ 1 + p_c\(\frac{\lambda}{M_u}\)^{\frac{\alpha}{2}}t \]^{-\frac{2}{\alpha-2}}\\
\emph{\text{Baseline}}: \quad \mathcal{P}_t (1) &\approx \[ 1 +
p_c\(\lambda\)^{\frac{\alpha}{2}}t \]^{-\frac{2}{\alpha-2}}
\end{align}\normalsize
\end{proposition}

The closed-form representations makes it possible to invert the reliability constraint equation $\mathcal{P}_t(M_i) = \eta$ with respect to $t$. Applying this $t$ to the rate $R_t(M_i)$ in \eqref{Eq:Rate} leads to the following closed-form ubiquitous rates.
\begin{proposition} \emph{(Ubiquitous Rate)} For $\eta \approx 1$, ubiquitous rates of $\bfrmath$, $\ufrmath$, and the baseline are given as follows.

\vspace{-10pt}
\small\begin{align}
\hspace{-8pt}\bfr: \mathcal{R}_{\eta}(M_b) &\approx  \frac{\eta W p_c\(\lambda\) \lambda_b }{M_b \lambda}   \log_{2}\(1 + \frac{c_\eta {M_b}^\frac{\alpha}{2} }{ p_c\( \lambda\)^\frac{\alpha}{2}}\)  \label{Eq:Prop2bfr}\\
\hspace{-8pt}\ufr: \mathcal{R}_{\eta}(M_u) &\approx  \frac{\eta W  p_c\(\frac{\lambda}{M_u}\) \lambda_b }{\lambda}   \log_{2}\(1 + \frac{c_\eta }{ p_c\( \frac{\lambda}{M_u}\)^\frac{\alpha}{2}}\) \label{Eq:Prop2ufr} \\
\hspace{-8pt}\emph{\text{Baseline}}: \mathcal{R}_{\eta}(1) &\approx  \frac{\eta W  p_c\(\lambda\)
\lambda_b }{\lambda}  \log_{2}\(1 + \frac{c_\eta }{ p_c\(\lambda\)^\frac{\alpha}{2}}\)
\end{align}\normalsize 
where $c_{\eta}:= \(\frac{\alpha}{2}-1\)(1-\eta)$

\begin{proof} See Appendix.
\end{proof}
\end{proposition}

This result specifies the impacts of frequency reuse and ultra-densification on ubiquitous rate as below.

\begin{remark} \emph{(Impact of $M_i$)} A larger $M_i$ decreases resource allocation, and yet increases spectral efficiency. Therefore, optimizing $M_i$ is required.
\end{remark}
\noindent For $\ufr$, the foregoing remark comes from the fact that $p_c(\lambda/M_u)$ is a monotone decreasing function of $M_u$.

\begin{remark} \emph{(Impact of Densification)} BS densification improves both spectral efficiency and resource allocation.
\end{remark}

By the definition $p_c(\lambda)$ decreases with $\lambda_b$, and asymptotically converges toward $0$ at the rate of $\lambda/\lambda_b$. This makes the density of the BSs occupying a single channel is upper bounded by the entire UE density, i.e. $p_c(\lambda)\lambda_b \leq \lambda$. Until reaching this bound, BS densification at least provides a marginal gain in resource allocation. Deploying more BSs also increases spectral efficiency by improving $\SIR$ reliability as specified in Lemma 3, leading to the above remark.

\begin{remark} \emph{(Indistinguishable Condition)} The ubiquitous rates of $\bfr$ and $\ufr$ are indistinguishable if $M_b$ and $M_u$ satisfy $p_c\(\lambda/M_u\)=p_c(\lambda)/M_b$, which is not always ensured since $M_u$ and $M_b$ are integers no smaller than 1.
\end{remark}
It straightforwardly follows from comparing \eqref{Eq:Prop2bfr} and \eqref{Eq:Prop2ufr}. This remark predicts that
the optimal ubiquitous rates of $\bfr$ and $\ufr$ are not always identical yet very close to each other, to be verified by simulation in the next section.

\begin{remark} \emph{(Identically Optimal Condition)} For the optimal $M_i^*$ maximizing its ubiquitous rate, the optimal ubiquitous rates $\mathcal{R}_\eta(M_b^*)$ and $\mathcal{R}_\eta(M_u^*)$ become identical when both $M_b^*,M_u^* \rightarrow 1$ or $\lambda_b \rightarrow \infty$.
\end{remark}
The case $M_i^*\rightarrow 1$ makes $\bfr$ and $\ufr$ converge to the baseline having $M_o=1$. For $\lambda_b\rightarrow \infty$, indistinguishable condition in Remark 3 is always satisfied since $p_c(\lambda)$ and $p_c(\lambda/M_u)$ converge to $0$.

\begin{figure}
	\centering
	\includegraphics[width=8.8cm]{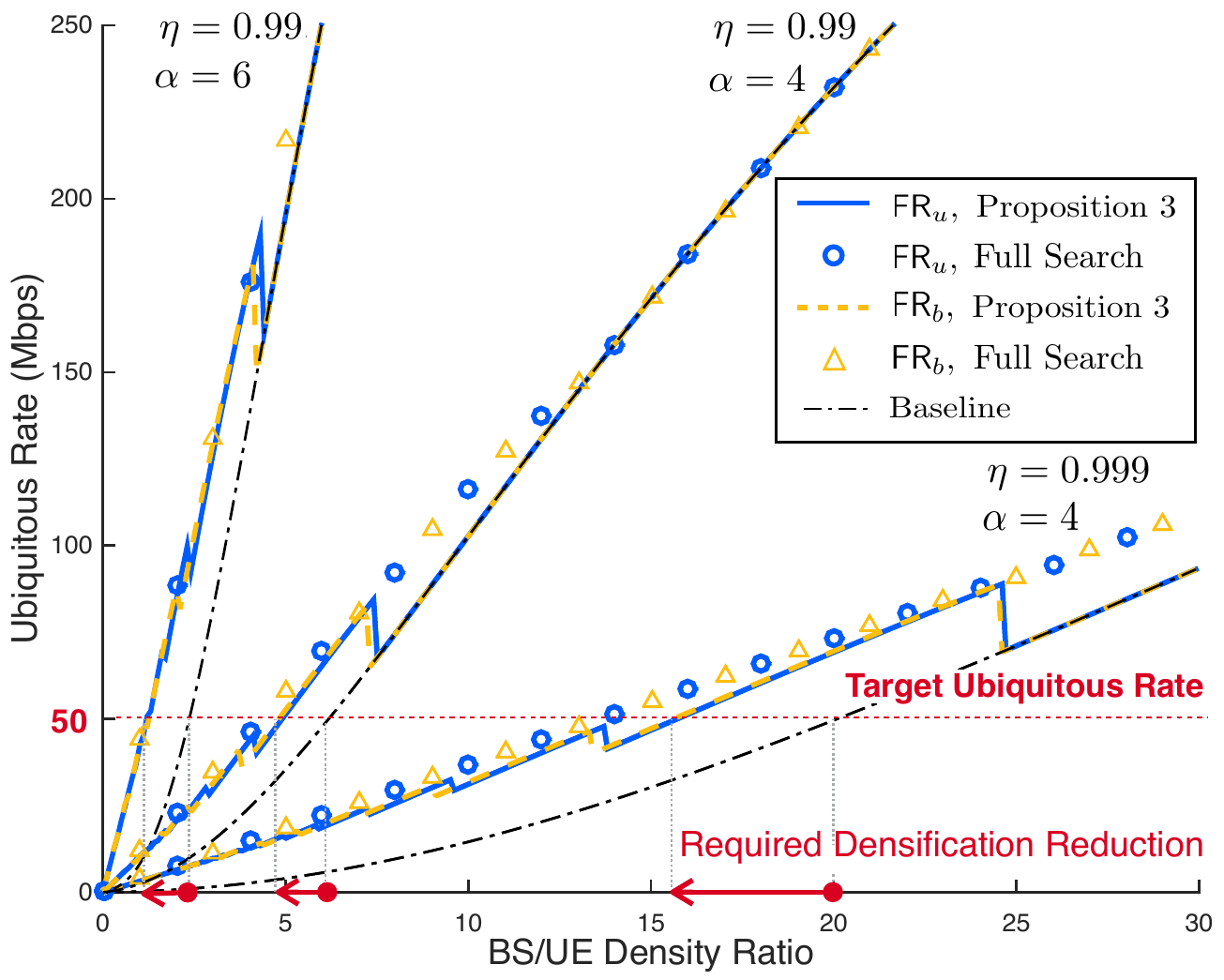}
	\caption{Maximized ubiquitous rates of $\bfr$ and $\ufr$ ($W = 100$ MHz). }
\end{figure}

\subsection{Optimal Frequency Reuse Design}
The optimal $M_i^*$ maximizing the ubiquitous rate is found as follows.

\begin{proposition} \emph{(Optimal $M_i^*$)} For $\eta \approx 1$, the optimal $M_i^*$ maximizing $\mathcal{R}_\eta(M_i)$ is given as below.

\vspace{-10pt}\small\begin{align}
\hspace{-15pt} M_b^* &\approx \underset{M_b}{\arg\min} \l| \(1+\frac{p_c\(\lambda\)^{\frac{\alpha}{2}}}{c_\eta {M_b}^\frac{\alpha}{2}}\)\log_2\(1+ \frac{c_\eta{M_b}^\frac{\alpha}{2}}{p_c\(\lambda\)^{\frac{\alpha}{2}}}\) - \frac{\alpha}{2}\r|  \\
\hspace{-15pt} M_u^* &\approx  \underset{M_u}{\arg\min} \l| \[  c_\eta + p_c\(\frac{\lambda}{M_u}\)^\frac{\alpha}{2} \] \log_2\(1 + \frac{c_\eta}{p_c\(\frac{\lambda}{M_u}\)^\frac{\alpha}{2}}\) - \frac{\alpha c_\eta}{2} \r|
\end{align}
\normalsize

\vspace{-5pt}\begin{proof} See Appendix.
\end{proof}
\end{proposition}

For $\lambda_b \rightarrow 0$ or $\infty$, the optimal $M_i^*$'s converge as follows.
\begin{corollary} \emph{(Asymptotically Optimal $M_i^*$)} For $\eta \approx 1$, the asymptotic behavior of the optimal $M_i^*$ with respect to BS density $\lambda_b$ is given as follows.

\vspace{-10pt}\begin{align}
M_b^*=M_u^* = \left\{\begin{array}{l l} \infty & {\text{if $(1-\eta){\lambda_b}^\frac{\alpha}{2}$}\rightarrow 0}  \\1 & {\text{if $(1-\eta){\lambda_b}^\frac{\alpha}{2} \rightarrow \infty$}}  \end{array}\right.
\end{align}\normalsize
\end{corollary}
 In the first diverging $M_i^*$, we can notice that the increase in $M_i^*$ comes from a higher reliability requirement $\eta$ and/or lower BS density. This implies exploiting frequency reuse becomes more effective for the UR2C design in sparse networks. The second converging $M_i^*$ to unity follows from the fact that BS ultra-densification can solely provide the target ultra-reliability while no longer relying on frequency reuse.

\begin{figure}
	\centering
	\includegraphics[width=8.8cm]{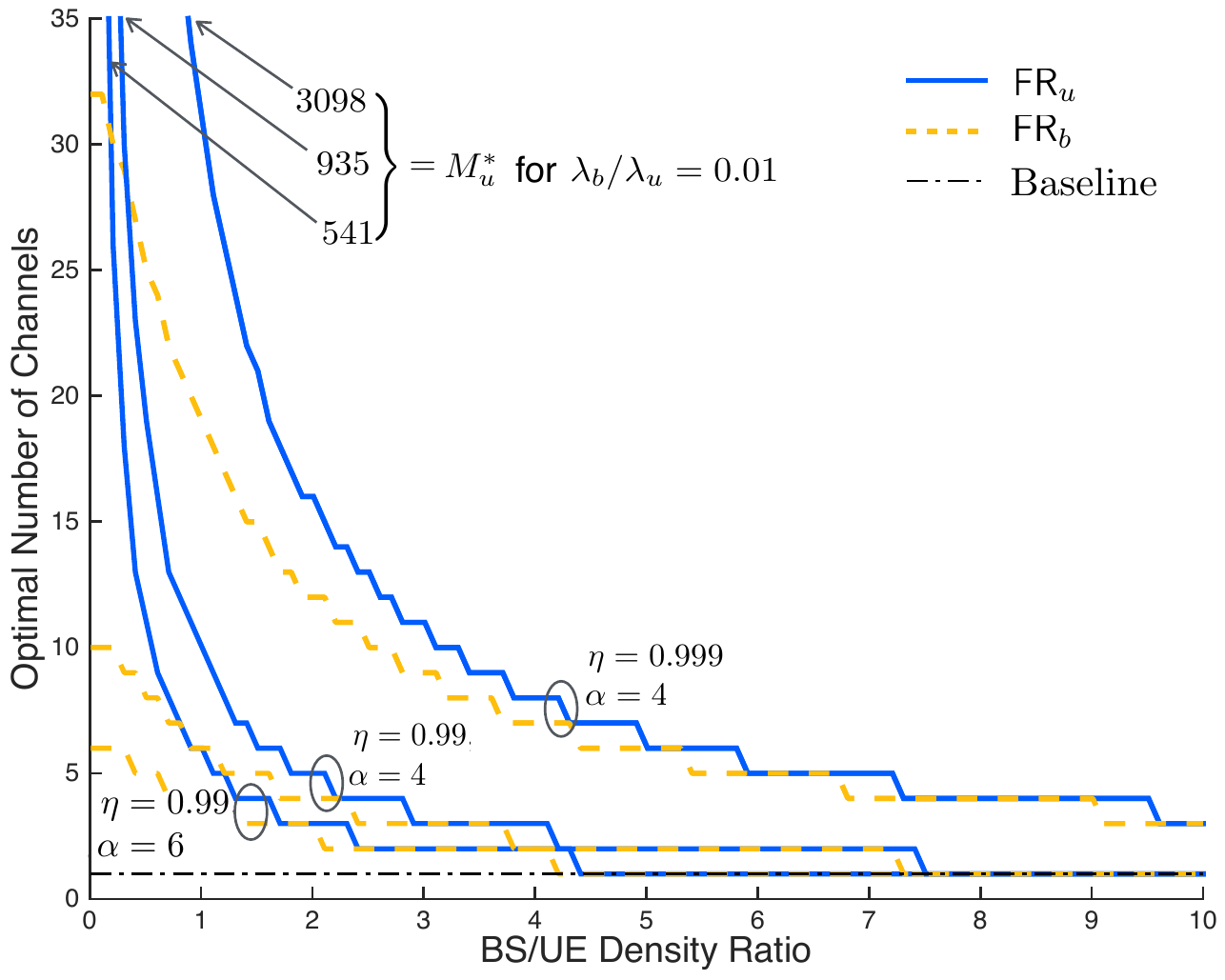}
	\caption{Maximized ubiquitous rates of $\bfr$ and $\ufr$ ($W = 100$ MHz). }
\end{figure}



\section{Numerical Evaluation}
Default system parameters in this section are given as follows: $\eta = 0.99$, $\alpha = 4$, and $W=100$ MHz \cite{Qualcomm5G:16}. Applying Proposition 3 to Proposition 2 via numerical simulation yields the following frequency reuse design guideline for UR2C.

\textbf{$\bfr$ and $\ufr$ increases ubiquitous rate, further improved by higher $\eta$ and $\alpha$}. Fig. 3 shows the maximized ubiquitous rates with $\bfr$ and $\ufr$ based on Proposition 3 (see yellow dotted and blue solid curves) outperform the baseline result without frequency reuse (black dash-dotted). The gain in ubiquitous rate, $\mathcal{R}_{\eta}(M_i^*)/\mathcal{R}_{\eta}(1)$, increases for higher $\eta$ as it makes frequency reuse more effective to cope with the target reliability requirement. Compared to the baseline, the maximum gain with $\eta=0.999$ is $3.15$ times larger than the gain with $\eta=0.99$. For higher $\alpha$, the gain also increases since $\SIR$ grows along with $\alpha$ under an interference-limited regime \cite{JHParkTWC:15,Andrews:2011bg,Haenggi:ISIT14}, amplifying the effect of frequency reuse in ubiquitous rate improvement. When $\alpha$ becomes $6$ from $4$, it increases the gain by $2.03$ times. For a given target ubiquitous rate $50$ Mbps, both $\bfr$ and $\ufr$ improve ubiquitous rate for all considered scenarios, while not converging to the baseline. This leads to reducing the required BS densification by up to $46\%$.

\textbf{$\bfr$ mostly performs better while $\ufr$ sometimes outperforms for large BS density}. Overall, the maximized ubiquitous rates of $\bfr$ and $\ufr$ keep crossing as BS density increases, but have no significant gap. As predicted in Remark 3, the reason follows from the fact that $M_i$'s are integers no smaller than $1$. These coarse unit changes cannot effectively minimize the objective argument functions in Proposition 3, so constantly lead to peaky crossings as well as the gap between the results from Proposition 3 and a full search algorithm in Fig. 3, until $\bfr$ and $\ufr$ converge to the baseline. While having similar ubiquitous rates, $\bfr$ requires less number of channels as shown in Fig. 4, so is more preferable than $\ufr$, especially for small BS density where $M_u^*$ is unrealistically large. For moderate BS density, $\ufr$ sometimes provides up to $28\%$ higher ubiquitous rate than $\bfr$, and it is thus important to compare them in this regime for UR2C design. For large BS density, frequency reuse gain in ubiquitous rate disappears as also anticipated in Remark 4 and Corollary 1.

\textbf{Optimal $\bfr$ and $\ufr$ provides similar per-UE resource allocations and spectral efficiencies}. Fig. 5 illustrates that $\bfr$'s maximum per-UE allocation $W/M_b^*$ is larger but its actual allocation fraction $\E[1/N_b^*]$ within the maximum value due to multiple access is smaller. These two factors cancelled out each other, resulting in almost the same resource allocation as $\ufr$'s. Fig. 6 visualizes $\bfr$'s maximized spectral efficiency keeps crossing but becomes clearly larger under a small BS/UE density environment. This originates from the fact that $\bfr$ more aggressively reduces interference by forcing to use only a single channel (see Fig. 1-a). It is effective under small BS/UE density where $\SIR$ is low, yielding $\bfr$'s smaller interferer density normalized by BS density $\lambda_i^*/\lambda_b$. Note also that non-smooth $\bfr$ and $\ufr$ curves in Figs. 5 and 6 are because $M_b$ and $M_u$ are integers as elaborated in Remark 3.

\begin{figure}
	\centering
	\includegraphics[width=8.8cm]{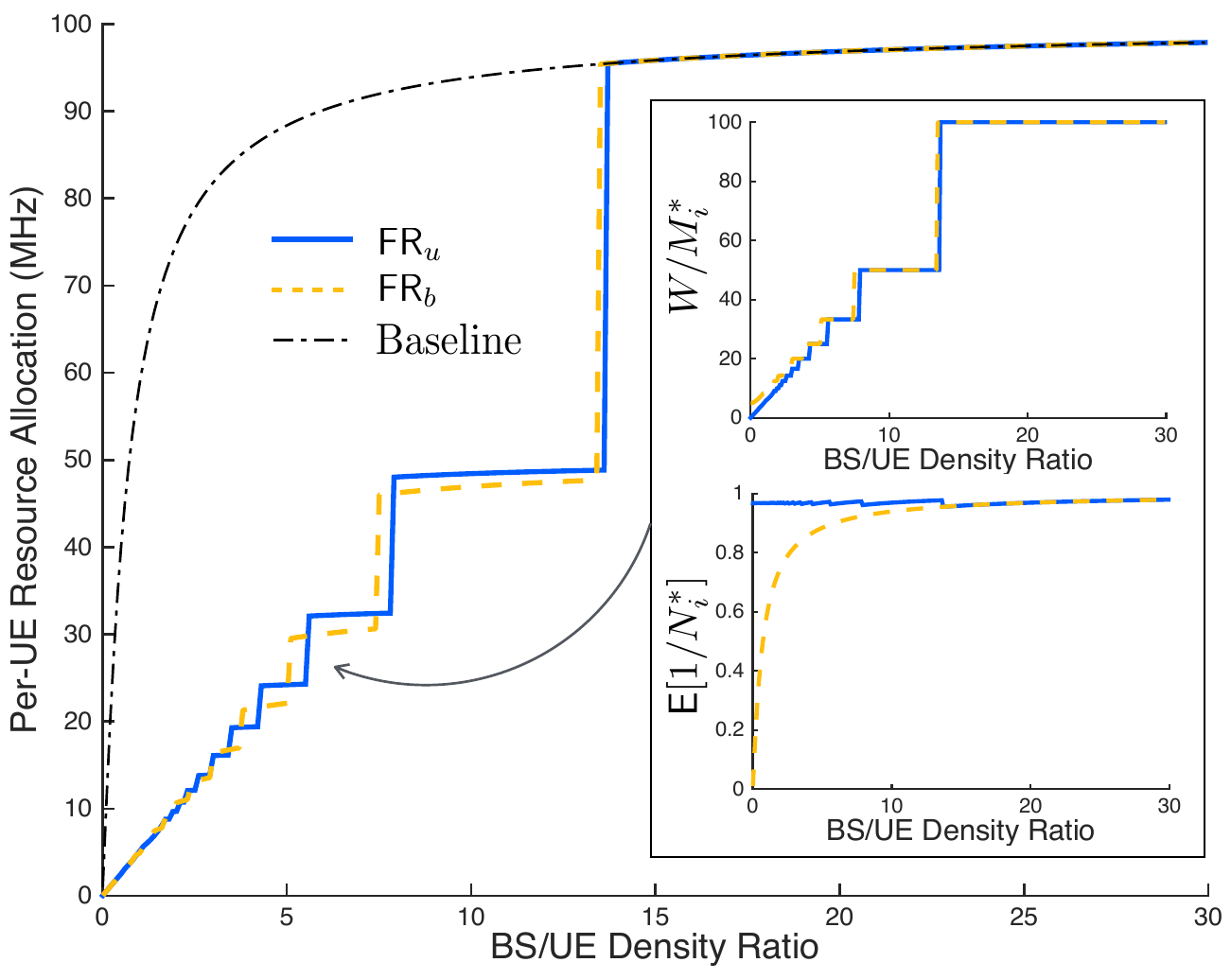}
	\caption{Maximized ubiquitous rates of $\bfr$ and $\ufr$ ($W = 100$ MHz, $\eta = 0.99$, $\alpha=4$). }
\end{figure}


\section{Conclusion}
In this paper we derive closed-form ubiquitous rate, and investigate two frequency reuse
schemes $\bfr$ and $\ufr$ so that they can ubiquitously achieve the UR2C's target $50$
Mbps data rate with $99\%$ $\SIR$ reliability. Both of their optimal designs provide
higher ubiquitous rates compared to the baseline without frequency reuse. This result is
different from traditional network design guidelines that prefer not to use frequency
reuse for only maximizing data rate without guaranteeing reliability. It would thus be worth revisiting more advanced frequency reuse techniques that additionally utilize spatial and power differences. To enable UR2C, it would also be an interesting topic to incorporate other reliability achieving schemes such as double associations for attaining diversity gain \cite{dmkim20016reliable}.

\begin{figure}
	\centering
	\includegraphics[width=8.8cm]{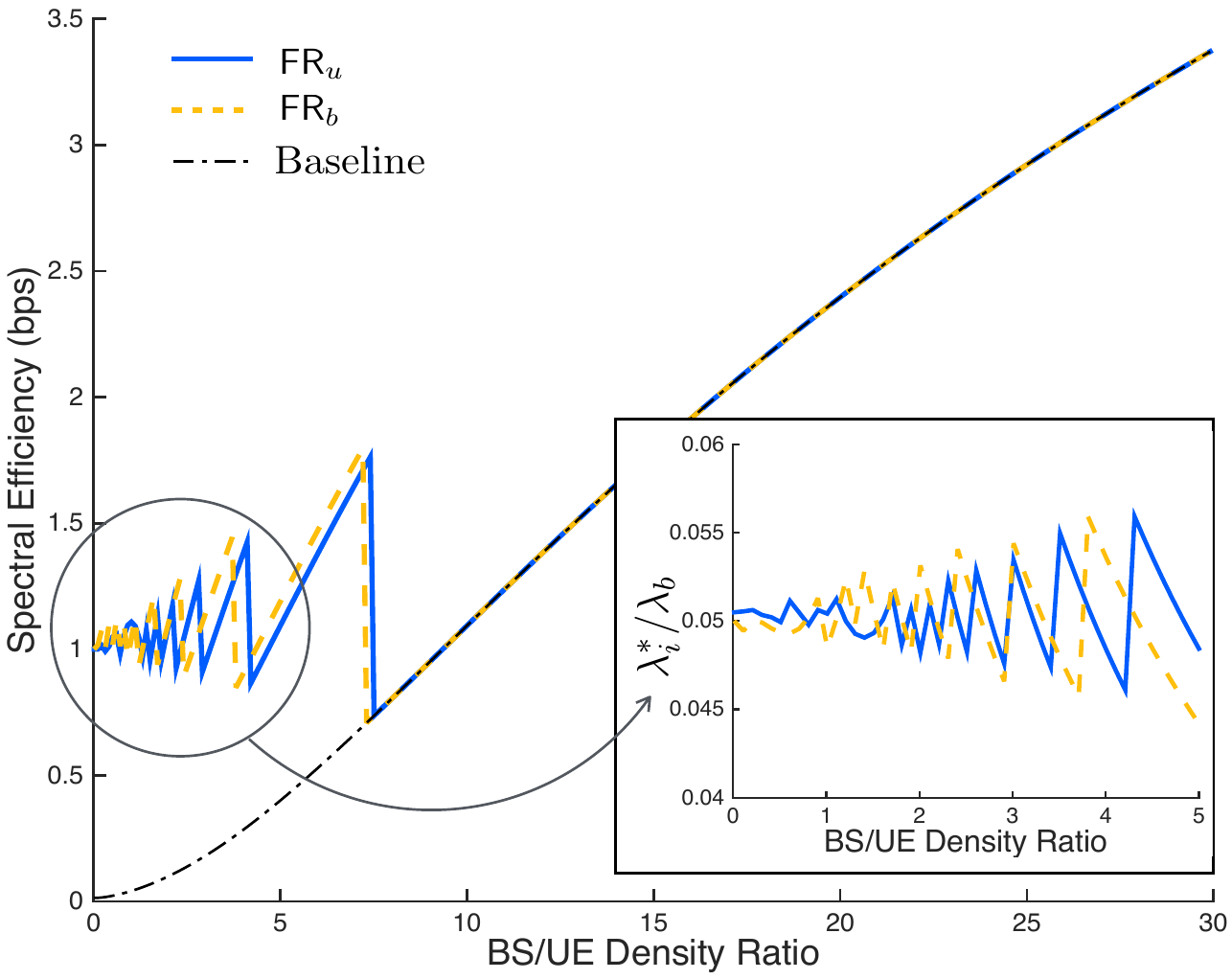}
	\caption{Maximized ubiquitous rates of $\bfr$ and $\ufr$ ($W = 100$ MHz, $\eta = 0.99$, $\alpha=4$). }
\end{figure}

\section*{Appendix}

\subsection{Proof of Lemma 3}
Let $f(x) = \;_2F_1\(1, -\frac{2}{\alpha}; 1-\frac{2}{\alpha}; -x\)^{-1}$. By the Gauss-hypergeometric function definition,

\vspace{-10pt}\small\begin{align}
\hspace{-3pt}f(x) &= \[\sum_{n=0}^\infty \frac{(1)_n \(-\frac{2}{\alpha}\)_n }{\(1-\frac{2}{\alpha} \)_n} \frac{(-x)^{n}}{n!} \]^{-1}= 1 - \frac{2 x}{\alpha-2} + O(x^2)
\end{align}\normalsize
The last step follows from Taylor's expansion for $x\rightarrow 0$. This indicates $\lim_{x\rightarrow 0} f'(x)=-2/(\alpha-2)$. 

Consider $g(x):=\( 1 + x  \)^{-\frac{2}{\alpha-2}}$, having the same slope as $f'(x)$ as $x\rightarrow 0$. To verify $g(x)\leq f(x)$, we consider the first and second derivatives of $f(x)$ and $g(x)$ for $x>0$ as follows.

\vspace{-10pt}\footnotesize\begin{align}
\hspace{-5pt} f'(x) &= -\frac{2\[ f(x) - \(1+x\)^{-1} f(x)^2  \]}{\alpha x} \\
\hspace{-5pt} g'(x) &= - \frac{2 (1+x)^{-\(1 + \frac{\alpha}{\alpha-2}\)}}{\alpha-2} \label{Eq:PfLemma3_g}\\
\hspace{-5pt}f''(x) &=  \frac{2(2+\alpha)f(x) + (1+x)^{-2}\[8-2( \alpha + 6 + 6x + 2 \alpha x)f(x)^{2}\]}{(a x )^2}\\
\hspace{-5pt}g''(x) &=  \frac{2 \alpha(1 + x)^{-2\(1 +\frac{2}{\alpha-2} \)}}{\(\alpha-2\)^2}
\end{align}\normalsize

Firstly, $f'(x), g'(x)<0$, $f''(x)>0$, and $g''(x)\geq 0$ for all $x>0$ where the
equality holds when $x\rightarrow\infty$ and/or $\alpha\rightarrow \infty$. Secondly,
$\lim_{x\rightarrow 0}f'(x)/g'(x)=1$ and $\lim_{x\rightarrow \infty}f'(x)/g'(x)\leq 1$
where the equality holds when $\alpha\rightarrow \infty$. Lastly, $f(0)/g(0)=1$ and
$\lim_{x\rightarrow\infty}f(x)/g(x)=1$ due to their original coverage probability
definitions. As $x$ increases, these conclude that $f(x)$ and $g(x)$ are monotonically decreasing, starting identically from unity while the decreasing slope of $g(x)$ is steeper than that of $f(x)$ due to $f'(x)/g'(x)\leq 1$ for all $x>0$. This concludes the proof of $f(x)\geq g(x)$ for all $x$. \endproof

\subsection{Proof of Proposition 2}
Let $a_i$  denote $p_c(\lambda_b)/M_b$, $p_c(\lambda/M_u)$, and $p_c(\lambda)$ respectively for $\bfr$, $\ufr$ and the baseline.
By applying Proposition 1,  the reliability constraint equation $\mathcal{P}_t(M_i)=\eta$ is rephrased with respect to $t$ as 
\begin{align}
t &= \frac{\eta^{-\(\frac{\alpha}{2}-1\)}-1}{{a_i}^\frac{\alpha}{2}}\\
&= \frac{\[1-(1-\eta)\]^{-\(\frac{\alpha}{2}-1\)}-1}{{a_i}^\frac{\alpha}{2}} \label{Eq:PfProp1}
\end{align}

In \eqref{Eq:PfProp1}, applying Taylor expansion for $\eta\approx 1$ makes $\[1-(1-\eta)\]^{-\(\frac{\alpha}{2}-1\)}$ become $1 + c_\eta$. Applying this with Table II and a typical UE's reliability $\eta$ to \eqref{Eq:Rate} provides the desired result.  \endproof

\subsection{Proof of Proposition 3}

According to Karush-Kuhn-Tucker (KKT) first and second order conditions, the local maximum of $\mathcal{R}_\eta(M_b)$ in \eqref{Eq:Prop2bfr} is the global maximum. Likewise, the local maximum of $\mathcal{R}_\eta(M_u)$ in \eqref{Eq:Prop2ufr} becomes its global maximum. Therefore, $M_b$ and $M_u$ can maximize $\mathcal{R}_\eta(M_b)$ and $\mathcal{R}_\eta(M_u)$ when respectively satisfying the first order conditions $d\mathcal{R}_\eta(M_b)/dM_b=0$ and $d\mathcal{R}_\eta(M_u)/dM_u=0$. Now that $M_b$ and $M_u$ are integers no smaller than $1$, the values minimizing the LHS of the first order conditions provide the optimal numbers of channels, finalizing the proof. \endproof

\bibliographystyle{ieeetr}  

\end{document}